\documentclass[letterpaper,twocolumn,10pt]{article}
\usepackage{usenix}

\usepackage[printonlyused]{acronym}
\usepackage[longend,ruled,linesnumbered]{algorithm2e}
\usepackage[american]{babel}		%
\usepackage{booktabs}			%
\usepackage[skip=1pt]{caption}		%
\usepackage{enumitem}			%
\usepackage{hyperref}			%
\usepackage[utf8]{inputenc}		%
\usepackage{subcaption}
\usepackage[dvipsnames]{xcolor}
\usepackage[nottdefault]{sourcecodepro}

\usepackage{listings}
\lstdefinelanguage{Toml}{
    comment = [l]{\#},
    keywords = {true, false},
    morestring = [b]{"}
}

\addto\extrasamerican{

}

\usepackage{multicol}			%
\usepackage{multirow}			%
\usepackage{nicematrix}       %
\usepackage{pgfplots}			%
\pgfplotsset{compat=1.18}
\usepackage{tikz}			%
\usepackage{todonotes}		%
\usepackage{xspace}			%
\usepackage{xurl}			%

\usetikzlibrary{patterns, shapes,positioning,graphs,quotes,calc}

\usepackage{cleveref}			%
\usepackage{fontawesome5}

\usepackage{amssymb}
\usepackage{amsmath}
\usepackage{pifont}
\usepackage{threeparttable}
\newcommand{\cmark}{\textcolor{green}{\ding{51}}}
\newcommand{\xmark}{\textcolor{red}{\ding{55}}}
\newcommand{\pmark}{\textcolor{orange}{$\boldsymbol{\sim}$}}

\lstdefinelanguage{sleigh}[]{C}{
  morekeywords={is, export, local, inst_next, inst_start, sext},
  emph={opsize0, opsize1, opsize2, opsize3, cond0, cond1, condX, end},
  basicstyle=\fontfamily{SourceCodePro-TLF}\selectfont\scriptsize,
  emphstyle=\color{violet},
  keywordstyle=\color{blue},
  stringstyle=\color{red},
  morecomment=[l]\#,
  commentstyle=\color{gray},
  numbers=left,
  numberstyle=\tiny\color{black},
  stepnumber=1,
  numbersep=5pt,
  backgroundcolor=\color{white},
  showspaces=false,
  showstringspaces=false,
  tabsize=2,
  breaklines=true,
  breakatwhitespace=false,
}

\lstdefinelanguage{reC}[]{C}{
  morekeywords={uint, byte, longlong, ulong, undefined8, undefined4},
  emph={DEKHash, binary, next_domain, fib},
  emph={[2],code_r0x0040618b},
  emphstyle=\color{red},
  emphstyle=[2]\color{violet},
  basicstyle=\fontfamily{SourceCodePro-TLF}\selectfont\scriptsize,
  keywordstyle=\color{blue},
  stringstyle=\color{red},
  commentstyle=\color{gray},
  numbers=left,
  numberstyle=\tiny\color{black},
  stepnumber=1,
  numbersep=5pt,
  backgroundcolor=\color{white},
  showspaces=false,
  showstringspaces=false,
  tabsize=2,
  breaklines=true,
  breakatwhitespace=false,
}

\lstdefinelanguage{pseudoAsm}[]{C}{
  morekeywords={push, ret},
  emph={jmptarget,f,g},
	emph={[2],},
  emphstyle=\color{violet},
  emphstyle=[2]\color{gray},
  basicstyle=\fontfamily{SourceCodePro-TLF}\selectfont\scriptsize,
  keywordstyle=\color{blue},
  stringstyle=\color{red},
  commentstyle=\color{gray},
  numbers=none,
  numberstyle=\tiny\color{black},
  stepnumber=1,
  numbersep=5pt,
  backgroundcolor=\color{white},
  showspaces=false,
  showstringspaces=false,
  tabsize=2,
  breaklines=true,
  breakatwhitespace=false,
}

\graphicspath{ {./} }

\newcommand{\RQ}[1]{\textbf{RQ{#1}}}
\newcommand{\apdxref}[1]{\hyperref[#1]{Appendix~\ref*{#1}}}

\newcommand{\toolname}{\mbox{VMSpec}}
\newcommand{\numprograms}{147\xspace}

\newcommand{\numdgas}{six\xspace}

\newcommand{\phead}[1]{\noindent\textbf{#1}}

\AtBeginDocument{

}

\begin{document}

\date{}

\title{\Large \bf\toolname: A Decompiler-based Approach against Virtual Machine Obfuscation}

\author{
	{\rm Fabian Kilger \qquad Leonard Auer \qquad Manuel Andreas}\\[0.7ex]
	Technical University of Munich
} %

\maketitle

\begin{abstract}
Malicious actors frequently obfuscate their code to hinder investigative
reverse engineering efforts. Among the most
formidable techniques is virtual machine (VM) obfuscation: programs
are translated into bytecode for a custom VM, often with
a randomized virtual instruction set and distributed alongside
a corresponding interpreter.

Manual analysis and deobfuscation scale poorly due to the
complexity of VM obfuscation and the diversification
techniques that render each VM and interpreter unique.
Existing methods primarily focus on simplifying execution traces
and identifying virtual execution boundaries. However, comparatively
little attention has been given to recovering higher-level
program semantics from VM-obfuscated code.
To address this gap, we propose \toolname{}, a novel decompiler-based
approach that automates the construction of VM processor
specifications, enabling existing, well-known decompilers to analyze the VM-obfuscated code.

We evaluate our prototype of {\toolname}
on \numprograms programs from an established obfuscation benchmark,
comprising hash functions, standard algorithms, and randomized functions,
obfuscated with different configurations of the Tigress and Code Virtualizer obfuscators.
To demonstrate real-world applicability, we
additionally include \numdgas domain generation algorithms stemming from real-world
malware.
\toolname{} completely strips VM-obfuscation from nearly all obfuscated samples,
producing output comparable to Ghidra’s decompilation of the
original code. Furthermore, it is the first method capable
of reconstructing the complete program behavior of VM-obfuscated programs
with infinite, input-dependent paths and recursive calls.

\end{abstract}

\section{Introduction}\label{section:introduction}

Program authors obfuscate their software to thwart reverse engineering
and other forms of program analysis.
While obfuscation can serve legitimate purposes -- such as protecting
intellectual property or reducing fraud -- it is especially attractive
for malware authors seeking to evade detection and analysis.
Among the various obfuscation techniques, virtualization-based
obfuscation is widely considered one of the strongest techniques available~\cite{sok}.
Here, the original program is translated into bytecode for a custom virtual machine (VM),
which is then embedded, together with a bytecode interpreter, into the obfuscated program.
The meaning of individual bytecode instructions is therefore encoded solely in the
interpreter, thereby forcing human analysts to first reverse-engineer
the interpreter before they can reason about the behavior of the
embedded bytecode.
Moreover, to hinder knowledge reuse, commercial obfuscators~\cite{tigress,themida,codevirtualizer,vmprotect}
generate distinct virtual machine architectures, instruction sets, and interpreters
by employing various diversification techniques~\cite{vmdoctor2025}.

Many techniques have been proposed to deobfuscate virtualization-based
obfuscated programs~\cite{sok} and obfuscated software in general~\cite{schrittwieser2016protecting}.
Current state-of-the-art methods against virtualization-based obfuscation primarily rely on
instruction traces collected during execution
of the obfuscated program. These methods analyze and simplify runtime traces either
directly~\cite{_vmhunt,_coogan,_tigress_deobf,_seead,_yadegari}
or first identify VM handlers, which then serve
as an anchor point for trace simplification~\cite{_vmattack,_syntia,_virtual_deobfuscator,_dynopvm,_liang}.
However, dynamic tracing approaches are inherently limited to the code path
taken during a given run, thereby limiting the analysis of program behavior
to the specific inputs provided.
While prior work~\cite{_tigress_deobf,_seead,_yadegari} proposes overcoming this problem using path exploration algorithms,
they cannot guarantee exhaustive path coverage and remain constrained by the path explosion problem,
which renders them infeasible for more complex programs.
Furthermore, studies~\cite{ollivier2019obfuscation,ollivier2019kill} show
that attackers can craft strong and robust primitives to explode
the number of paths. Consequently, dynamic tracing approaches
are unable to recover the complete behavior of virtualization-based
obfuscated programs.

Towards automated extraction of the complete program behavior,
prior works proposed methods for identifying
state variables~\cite{_rotalume}, VM boundaries~\cite{_vmhunt}, and
handler boundaries~\cite{_vmhunt,_syntia,_rotalume}.
To extract handler behavior, symbolic execution combined
with compiler optimizations~\cite{_liang} and program synthesis~\cite{_syntia,xyntia}
has been proposed. However, the extracted handler behavior is incomplete
since the analysis is restricted to single handler traces and ignores control flow.

To address these shortcomings, we propose \toolname{},
a novel method that automates the generation of SLEIGH processor specifications~\cite{sleigh}
for unknown VMs, thereby enabling analysts to leverage
existing decompilers and analysis frameworks.
As such, \toolname{} can produce a semantically complete decompilation
of the obfuscated program.

Given a description of the VM state, we apply dynamic symbolic
execution (DSE) to recover the behavior and encoding of VM instructions.
To enable complete and efficient analysis of handler behavior
in the presence of symbolic addresses, we introduce a custom memory model.
While DSE is susceptible to path explosion
and complex constraints, bytecode handlers contain
relatively little logic, making these problems simpler to mitigate.

We evaluate the semantic correctness and quality of our decompilation
output against a benchmark of \numprograms programs consisting of
hash functions, standardized algorithms, and randomized functions
used in prior work~\cite{sok,vmdoctor2025,banescu2016samples}.
To demonstrate \toolname{}'s real-world applicability, we further include
six command-and-control server domain generation algorithms (DGAs)
extracted from real-world malware.

Evaluating several configurations of the Tigress~\cite{tigress} obfuscator and Code Virtualizer~\cite{codevirtualizer} obfuscator, we find that \toolname{}
recovers decompiled code that is semantically correct and comparable to the decompilation of the unobfuscated
samples. In particular, \toolname{} is the first approach to demonstrate complete reconstruction of program behavior
on samples with more complicated control flow, such as input-dependent loops that typically lead to path explosion.

In summary, we make the following main contributions:

\begin{itemize}
	\setlength\itemsep{0pt}
	\setlength\parsep{0pt}
	\item We present \toolname{}, a decompiler-based approach against VM obfuscation,
		 that applies DSE to automate the analysis of bytecode handlers and generates instruction encodings and
		 behavior in the SLEIGH language.
	\item  \toolname{} enables existing decompilers and analysis frameworks to
			recover high-level program behavior from VM-obfuscated binaries.

	\item \toolname{} is the first approach that can reconstruct
	      the complete behavior of VM-obfuscated programs with
	      infinite, input-dependent paths and recursive calls.
    \item A prototype of \toolname{}, which we open-source upon publication in conference proceedings.

\end{itemize}

\section{Background}\label{section:background}

In this section, we first provide an overview of VM
obfuscation and existing deobfuscation approaches. Next, we explain how
the SLEIGH processor specification language~\cite{sleigh} allows
extending Ghidra and other analysis frameworks.

\subsection{Virtual Machine Obfuscation}\label{subsection:vm_obfuscation}

Virtual machine obfuscation is also referred to as
virtualization-based obfuscation or simply virtualization.
Despite its name, it is not related to VMs in the
context of hypervisors. VM obfuscation involves transforming
code into bytecode executed by an interpreter, the VM.

In the context of obfuscation, virtual machines emulate a
custom processor architecture. Virtualization obfuscators
first generate a \textit{virtual instruction set} that defines
the instructions in the VM bytecode.
During obfuscation, the original code is then translated to bytecode,
and the \textit{bytecode interpreter} is packaged into the final
binary~\cite{sok}.

Borrowed from Salwan et al.~\cite{_tigress_deobf}, we can describe
the interpreter as a loop that performs the following five steps:

\begin{enumerate}
	\setlength\itemsep{0pt}
	\setlength\parsep{0pt}
	\item The interpreter \textit{fetches} the next virtual instruction
		   from the \textit{virtual program counter} (VPC).

	\item The second step \textit{decodes} the virtual instruction:
	      It determines the opcode and operands.

	\item Then the interpreter selects the VM handler matching
	      the opcode, and \textit{dispatches} execution to that handler.

	\item The \textit{handler} executes the virtual instruction.

	\item Last, the \textit{terminator} returns the execution flow
	      from the handler to the interpreter loop.
\end{enumerate}

To prevent knowledge reuse between samples, obfuscators
employ a wide variety of diversification techniques~\cite{vmdoctor2025}.
Commercial obfuscators append the fetch, decode, dispatch,
and terminator steps to the end of each handler~\cite{_vmhunt} and
encrypt operands~\cite{vmdoctor2025}.
Furthermore, unlike in standard processor designs, bytecode may not
be contiguous, and each instruction then points to the location
of the next instruction.

\subsection{VM Deobfuscation Approaches}

In this section, we collect existing approaches
for extracting semantics from VM-obfuscated code and VM handler
analysis, and identify key techniques and their limitations. Then,
we briefly discuss how \toolname{} overcomes those limitations, motivating our design.
We provide an overview of deobfuscation approaches in \autoref{table:related-work} and
an overview of handler analysis approaches in \autoref{table:related-work-handler}.

\begin{table*}
\centering
\resizebox{\linewidth}{!}{%
\begin{threeparttable}
	\centering
    \setlength{\tabcolsep}{3pt}
	 \caption{Comparison of \toolname{} with existing approaches for VM-deobfuscation.}
	\label{table:related-work}
	\begin{tabular}{l l l c c c c c}
	\toprule
	Approach                       & Key Techniques                                   & Outcomes                                   & Requires identified & Resistant to     & Complex Programs & Decompiled Code & Complete Semantics \\
                                  &                                                  &                                            & VM state            & Diversification  &                  &                 &                    \\
	\midrule
        Rolles~\cite{_rolles}     & Pattern Matching, Lifting, Compiler Opt.         & Devirt. Code                               & \cmark              & \xmark \tnote{1} & \xmark & \cmark & \xmark \\
        Rotalumè~\cite{_rotalume} & Trace, Taint Analysis                            & Trace CFG, Bytecode Decoding               & \xmark              & \xmark \tnote{1} & \cmark & \xmark & \xmark \\ 
        Coogan~\cite{_coogan}     & Trace, Slicing                                   & Subtrace                                   & \xmark              & \xmark \tnote{1} & \cmark & \xmark & \xmark \\
        Kinder~\cite{_kinder}     & VPC-sensitive Static Analysis                    & External calls \& arguments                & \cmark              & \xmark \tnote{1} & ? \tnote{2} & \xmark & \xmark \\
        Raber~\cite{_virtual_deobfuscator} & Trace, Pattern Matching                 & Simplified Trace                           & \xmark              & \xmark           & \cmark & \xmark & \xmark \\
        Yadegari~\cite{_yadegari} & Trace, Taint Analysis, Concolic Exec.            & Trace, CFG                                 & \xmark              & \xmark \tnote{1} & \cmark & \xmark & \xmark \\
        SEEAD~\cite{_seead}       & Trace, Trace steering                            & Simplified Trace, CFG, FCG                 & \xmark              & \xmark \tnote{1} & \xmark & \xmark & \xmark \\
        VMHunt~\cite{_vmhunt}     & Trace, Symbolic execution                        & Simplified Trace, Semantic Formula         & \xmark              & \cmark \tnote{3} & \cmark & \xmark & \xmark\\
        Liang~\cite{_liang}       & Trace, Symbolic execution, Compiler Opt.         & Simplified Trace                           & \xmark              & ?                & ? \tnote{2} & \cmark & \xmark\\
        Salwan~\cite{_tigress_deobf} & Trace, Symbolic execution                     & Recompiled Code                            & \xmark              & \cmark           & \xmark & \cmark & \xmark\\
        PUSHAN~\cite{_pushan}     & VPC-sensitive Symbolic execution                 & CFG, Decompiled Code                       & \cmark              & \cmark           & \cmark & \cmark & \pmark \\
        VMSpec                    & Symbolic execution, Static Analysis (Decompiler) & Processor Spec., Decompiled Code           & \cmark              & \cmark           & \cmark & \cmark & \cmark \\
	\bottomrule
	\end{tabular}
    \begin{tablenotes}

    \item[]\makebox[\linewidth][c]{%
    \textsuperscript{1} Analysis taken from \cite{vmdoctor2025}.\quad
    \textsuperscript{2} Only evaluated on simple programs.\quad
    \textsuperscript{3} Runtime enhanced by \cite{vmdoctor2025}.%
    }
    \end{tablenotes}
\end{threeparttable}
    }
\end{table*}

\begin{table}
\centering
\resizebox{\linewidth}{!}{%
\begin{threeparttable}
	\centering
    \setlength{\tabcolsep}{3pt}
	\caption{Existing Approaches for automated VM-Handler Analysis.}
	\label{table:related-work-handler}
	\begin{tabular}{l l l c c c c}
	\toprule
	Approach                       & Key Technique                                & Scope                     & Control-Flow          & Memory Semantics  & Complete     \\
	\midrule
        Syntia~\cite{_syntia}     & Program Synthesis                            & Trace                     & \xmark                & \xmark            & \xmark       \\
        Xyntia~\cite{_syntia}     & Program Synthesis                            & Trace                     & \xmark                & \xmark            & \xmark       \\
        Liang~\cite{_liang}       & Symbolic execution                           & Trace                     & \xmark                & \pmark\tnote{1}   & \xmark       \\
        VMSpec                    & Symbolic execution                           & Full Analysis             & \cmark                & \cmark            & \cmark       \\
	\bottomrule
	\end{tabular}
    \begin{tablenotes}
    \item[1] Does not consider aliasing. For example, assumes ptr+variable and ptr+1 never alias.
    \end{tablenotes}
\end{threeparttable}
    }
\end{table}

\phead{Trace-based Deobfuscation.}
The vast majority of methods collect instruction traces from obfuscated samples
and apply various static and dynamic analysis techniques to simplify
the runtime traces. Their advantage is that these approaches
do not rely on identifying characteristics of the emulated VM.
Several works reconstructed the original program
by analyzing control and data dependencies~\cite{_coogan,_seead,_yadegari};
however, a recent study~\cite{vmdoctor2025} suggests that
these approaches are no longer applicable to modern diversification
techniques. Other approaches~\cite{_vmhunt,_tigress_deobf,_liang} employ
DSE to simplify runtime traces. VMHunt~\cite{_vmhunt} first identifies
the start and end of virtualized code in a runtime trace and
performs DSE only on the virtualized parts of the trace.
Salwan et al.~\cite{_tigress_deobf} apply DSE to the entire program
to identify input-output relations and directly translate them to C code.
Liang et al.'s~\cite{_liang} DSE approach differs in that
they analyze single handler traces and generate C code for each 
handler trace based on DSE results. Then they create a sequence
of calls representing each handler trace and use a compiler for simplification.
All of these methods, however, cannot extract the complete program
semantics because they are restricted to analyzing runtime traces.
Some approaches~\cite{_tigress_deobf,_seead,_yadegari} attempt
to overcome this restriction by including multiple traces:
They employ path exploration techniques to generate numerous inputs
and merge the resulting traces to a control flow graph, enabling
the recovery of loops. However, recent studies~\cite{ollivier2019obfuscation,ollivier2019kill}
present strong and robust primitives that thwart these
approaches by exploding the number of possible paths at no cost.

\phead{VPC-sensitive analysis.}
Contrary to trace-based approaches, some works~\cite{_rolles,_kinder,_pushan}
propose \textit{VPC}-sensitive analyses. While such analyses
require prior reverse-engineering of the VM, they allow reasoning over
the semantics of the \textit{whole} program. Rolles~\cite{_rolles}
manually built a disassembler and lifter and adapted it to other
samples using pattern matching. Equipped with a bytecode lifter,
reasoning about the complete obfuscated program becomes possible.
However, with modern diversification techniques, pattern-based
adaptations are no longer possible~\cite{vmdoctor2025}.
Kinder~\cite{_kinder} describes how
static analysis can be enhanced with VPC-sensitivity and prototype
an example value analysis. However, they evaluate their approach
only on the Fibonacci function. PUSHAN~\cite{_pushan} employs a \textit{constraint-free}
symbolic execution approach over the \textit{whole} program. This
approach allows them to recover a complete control-flow graph (CFG)
of the program. They also extend a decompiler to process their
extracted CFG. While they show reliable semantic recovery for
stack-based VMs on randomized hash functions, semantic correctness
has not been shown for more complex programs. Additionally,
a crucial limitation we identify is the unsound memory
model: symbolic pointer indexing is always resolved to \textit{one}
specific index, potentially under-approximating the control flow. In
particular, \textit{jump tables} produce this type of control flow,
which cannot be recovered with this memory model.
\toolname{} employs a VPC-sensitive analysis by producing
a processor specification that decompilers can consume,
treating the VPC as the program counter. Compared to PUSHAN,
\toolname{} leverages existing CFG recovery techniques
implemented in decompilers.

\phead{Handler Analysis.}
To analyze the semantics of handlers, program synthesis and
symbolic execution have been proposed. All existing approaches
analyze only \textit{handler traces}, i.e., extract the semantics of
a single path through the handler. Consequently, the extracted
semantics are incomplete. Two works~\cite{_syntia,xyntia}
successfully apply program synthesis to simplify expressions in VM
handlers and to identify arithmetic handlers. However, existing
program-synthesis-based deobfuscation approaches cannot recover
control flow nor memory semantics. Liang et al.~\cite{_liang}
employ symbolic execution to extract the semantics of
a handler trace. Notably, they extract semantics in the context
of the \textit{native} processor, not the VM semantics.
Furthermore, control-flow information is neither
considered nor extracted. In terms of memory semantics, they
employ Miasm~\cite{miasm}, which assumes a simplified memory
model in which an expression \texttt{ptr+var} will never alias
with \texttt{ptr+1}~\cite{miasm_memory}. Overall, existing
handler analysis approaches are insufficient to generate
complete instruction semantics appropriate for a VM
processor specification. In comparison, \toolname{} extracts
instruction encodings and VM semantics, including control-flow,
from the complete bytecode handler.

\subsection{Ghidra and SLEIGH}\label{subsection:sleigh}

Ghidra~\cite{ghidra} is an open-source reverse engineering
suite developed by the National Security Agency. It includes a decompiler that
produces C-like pseudocode. Ghidra uses an intermediate representation (IR)
called \mbox{P-Code} to reuse decompiler analyses. To specify the underlying
processor architecture and how instructions map to \mbox{P-Code},
the Ghidra developers designed the SLEIGH processor
specification language~\cite{sleigh}. In this language, engineers specify the
instructions' byte encoding and \mbox{P-Code} semantics.

\begin{figure}

	\begin{lstlisting}[language=sleigh, numbers=none,xleftmargin=1.2mm,xrightmargin=0.8mm]
cc: "O"  is cond=0  { export OF; }
 ...
cc: "Z"  is cond=4  { export ZF; }
cc: "NZ" is cond=5  { local tmp = !ZF; export tmp; }
 ...
rel32: reloc is simm32 [ reloc=inst_next+simm32; ] {
	export *[ram]:8 reloc;
}
:J^cc rel32 is opsize=2 & byte=0xf & cc; rel32 {
	if (cc) goto rel32;
}
	\end{lstlisting}
	\caption{Simplified SLEIGH instruction specification for an x86/64
	         conditional jump instruction as implemented in Ghidra.}
	\label{figure:cjmp_example}
\end{figure}

We present a simplified version of Ghidra's SLEIGH implementation of the x86/64 conditional jump instruction
in \autoref{figure:cjmp_example}. The basic syntactic
unit for instruction definitions is the \textit{constructor}. It consists
of a symbol name followed by the display, bit pattern, disassembly expression,
and semantic sections. There can be multiple constructors for each
symbol, identified by different bit patterns, as seen with \texttt{cc}.
An empty symbol name is used for instructions.

The display section defines the disassembler output, which consists of
literals, operands, or expressions defined in the disassembly
expressions section enclosed in square brackets. If referenced symbols define
their own disassembly output, it can be concatenated with the \texttt{\^}
operator.
After the \texttt{is} identifier, the bit pattern section follows, which
defines matching rules for a constructor. The statements
in the curly brackets refer to the semantics section, which
defines the P-Code semantics.
Export statements define the result used if
a symbol is referenced in another constructor. In the case of
\texttt{cc}, this is used to specify different expressions
depending on the value of \texttt{cond}.
Lastly, the symbol \texttt{rel32} is used to compute the absolute address of
a relative jump. It computes the expression for \texttt{reloc} inside
the disassembly expression section.

From an optimization perspective,
disassembly expressions provide an efficient way to specify
values that can be constant-propagated before \mbox{P-Code} emission. Furthermore,
branching decisions that can be resolved based on operands are
typically implemented with bit-pattern matching, avoiding dead branches
on operand values in the emitted \mbox{P-Code}.

\section{Virtual Machine Model}\label{subsection:model}

Before presenting \toolname{}, we define a general VM model
that specifies the individual components of a VM.
This model creates the basis for analyzing and
defining handler behavior. In the following, we explain
the components of our model.

\phead{Code.} This region contains the bytecode executed by
		the VM. The virtual program counter (VPC) points to the current
		instruction; in our model, it is a register called \texttt{VPC}.

\phead{Stack.} Most VMs use a stack pointer to point to the
		top of the stack; we model it as the \texttt{VSP} register.
		This register may be unused if a VM does not use a stack.

\phead{VM State Regions.} VMs may operate on internal state regions
		such as registers or data arrays that store local variables.

\phead{VM Context.} The instruction-decoding logic may depend
		on context information from previous instructions, which must be tracked for
		proper disassembly. For example, the opcode of an instruction
		may be located in the previous instruction. In this case,
		the stored opcode acts as context
		for decoding the next instruction. Furthermore,
		obfuscators may encrypt the opcode and operands; in this case, we also consider
		all variables involved in the encryption to be relevant context.

\phead{Global data.} The virtualized code typically includes
		references to regions outside the previously defined components. In
		particular, programs read and modify global data,
		including strings. In our model, we treat these as
		absolute references.

\section{Design}\label{section:design}

The primary goal of \toolname{} is to enhance a reverse engineer's
ability to understand VM-obfuscated code.
To achieve this, \toolname{} considers the VM an unknown processor architecture
and infers a SLEIGH processor specification by analyzing
the VM interpreter. The specification describes both a
\textit{disassembler} and a \textit{lifter} for the VM, thereby enabling
further downstream analysis, particularly decompilation.

\input{./design_flow_new.tex}

We depict an overview of \toolname{} in \autoref{figure:design_flow}.
First, preprocessing extracts general information
about the virtual machine, such as the location of the VM state (\textcolor{green!60!black}{\ding{192}}).
As a preprocessor may be unable to directly infer all opcodes and their
respective handler boundaries, \toolname{} includes an extensible
handler detector module (\textcolor{green!60!black}{\ding{193}}) that can query results of the handler
analysis to detect new handlers. The handler detector module then
informs the handler analysis step about the code boundaries of the respective handler.
This iterative process between the handler detector and analysis
is repeated until no more handlers can be discovered.

To extract handler semantics, we symbolically execute all paths
of a single handler (\textcolor{green!60!black}{\ding{194}}). As each conditional branch in the dynamic
symbolic execution forks an execution state, the primary result of this
analysis is a \textit{state tree} that tracks all states, including all
relevant events between states. Then, we apply static program
optimizations and merge duplicate nodes in the state tree,
resulting in a \textit{state graph} (\textcolor{green!60!black}{\ding{195}}).
Afterward, \toolname{} translates the state graph to code in a
processor specification language (\textcolor{green!60!black}{\ding{196}}). Finally, the processor
specification is fed to a decompiler to produce deobfuscated program code (\textcolor{green!60!black}{\ding{197}}).
In the following, we describe each step in more detail.

\subsection{Preprocessing and Handler Detection}\label{subsection:initial_analysis}

Preprocessing extracts information about the VM state and bytecode.
Then, the handler detection module discovers handler code boundaries
and forwards them to our DSE analysis.

Prior work~\cite{vmdoctor2025,_rotalume} proposed generic
approaches to detect this information in an automated way. However,
no implementations are currently available.
As we consider re-implementing generic solutions out of scope, our prototype
of \toolname{} implements specific solutions for the Tigress and Code Virtualizer
obfuscators, as detailed in \autoref{section:impl}.

\subsection{Symbolic Execution of VM Handlers}\label{subsection:handler_analysis}

We now describe the details of our handler analysis and
how we leverage DSE to extract semantics for each handler.
Overall, we use DSE to systematically explore all feasible execution paths
and record all memory operations (including pointer and value expressions)
and function calls. Based on this recorded information, we detect handler
operands and generate a state graph used in the code generation step.
In the following, we describe our DSE design in more detail.

\phead{Memory Model.}
A common problem for DSE are accesses through (unconstrained) symbolic pointers.
By default, DSE engines concretize these pointers to a subset of possible values
and fork the exploration for each concrete pointer value. This
inevitably results in incomplete semantics for unconstrained
pointers (such as current register values). However, when extracting the semantics of a
handler, we generally do not need to precisely resolve
all memory reads and writes. Instead of concretizing the pointers, it is sufficient to record
the occurrence of memory operations without introducing incorrect
assumptions about their values. Therefore, we create a new
symbolic variable for each read operation, except when the read
involves addresses in read-only memory.

Additionally, we cache recently written values to avoid creating new
variables when the program reads from
an address that has been modified recently. For this purpose, we adopt
a \textit{segmented} memory model based on the regions specified
in the VM state definition. This design allows us to retain the
cache as long as the symbolic address is restricted to a single
memory region, avoiding unnecessary cache flushes.

\phead{Function Calls.}
Calls change the control flow of the handler and execute external code.
During DSE, we generally do not follow function calls to external
code or other known functions; instead, we record the call site
and target, and model the call in the code generation step.
This allows us to significantly improve performance
as the number of paths and the complexity of potential constraints
are significantly reduced. The resulting code
will still contain a function call and can be further analyzed by
other tools or a reverse engineer.

\phead{Specialized semantics.}
Some handler semantics are incomplete without considering the concrete
values of operands and context. For example, binary-based virtual machine obfuscators
perform a native context switch to execute function calls or native instructions
that have not yet been implemented~\cite{li2022chosen}. The target location
may be encoded as an encrypted operand or depend on prior instructions.
Thus, the semantics cannot be inferred solely from the handler code.
We detect this case by identifying whether the native instruction pointer
after symbolic execution refers to neither a dispatcher nor another handler.
In that case, we extract specialized semantics by running DSE on the handler code,
using the concrete context and operands in the bytecode. In the following,
we refer to this technique as \textit{context concretization}.

Additionally, we use concretization to mitigate path explosion.
After concretizing the context and operands, conditions involving only these will
yield a concrete value,
therefore significantly reducing the number of explored paths. In particular,
we employ context concretization once the number of active states exceeds 100.

\phead{State Graph Generation.}
After DSE finishes, we obtain
a \textit{state tree} with edges representing conditional state transitions
and nodes containing the list of recorded calls and memory accesses.
To simplify the extracted semantics, we first remove dead memory
operations. Specifically, we perform a def-use analysis to identify
and remove dead memory stores and memory read operations whose
result is not used by the handler. Since we create a new variable for each memory read,
we can directly infer whether a variable is used without
performing complex data-flow analysis on memory locations.
Lastly, we merge nodes with the same content and successors to reduce duplicate code,
resulting in the final \textit{state graph}.

\phead{Handler Operand Identification.}\label{subsection:operand_discovery}
For automated identification of operands and their sizes, we employ
the heuristic proposed by Sharif et al.~\cite{_rotalume}, which
assumes operands are located at a small, constant offset from
their corresponding opcode. Accordingly, memory reads
near the VPC indicate operand accesses. However, code obfuscators
also include several dead VPC-relative memory accesses outside the
actual instruction. Therefore, we only record these VPC-relative
accesses during DSE and apply the heuristic after the
simplifications in the state graph generation step.

\subsection{Code Generation}\label{subsection:codegen}

Using the VM state definition, state graph, and detected operands,
we create a processor specification for the VM. The details
of this step depend on the concrete processor specification language.
In general, we need to specify the encoding (i.e., opcode and operands)
and semantics of each instruction. Overall, our code generation follows a standard
procedure: we iterate over the state graph in topological order and
translate each statement directly into its equivalent in the processor specification language.
We explain the details for SLEIGH in more detail in
\autoref{subsection:sleigh_codegen}.

\subsection{Decompilation and Postprocessing}

Finally, we package the generated processor specification into a
Ghidra plugin and use it to decompile the virtualized code.
We load the obfuscated binary into Ghidra using our custom
processor, allowing inspection of references to global data and variables.
To produce readable, correct code, we disassemble
the bytecode and specify the calling convention to the target
and all referenced functions.

\section{Implementation}\label{section:impl}
We implement the handler analysis of \toolname{} on top of the \textit{angr}~\cite{angr} framework.
We use hooks to record calls, memory accesses, and conditional jumps,
as well as to implement our custom memory model. Overall, the general parts are
implemented in 6883 Python lines of code (LOC) and 1296
Java LOC for Ghidra analysis code. Furthermore, we fixed several Ghidra bugs,
improved the soundness of Ghidra's stack analysis, and added a rule that triggers Ghidra's
SIMD-splitting routine on operations involving operands larger than 8 bytes. In total, our changes
touch 83 Java and 1062 C++ LOC. %

In the following, we describe our obfuscator-specific implementation details, the code generation
for SLEIGH, as well as our approach to recompiling Ghidra's decompilation output.

\subsection{Context Analysis}

To identify the necessary context for analyzing specialized handler semantics,
we apply recursive-sweep disassembly
while tracking the current VM context for each virtual program
counter. Our current prototype only analyzes \textit{local} context,
i.e., the state of VM encryption registers, but does not track
\textit{program} context (stack state, register state).
If \textit{local} context is insufficient, our implementation will
output a custom P-code operation that specifies the target
address of an unresolved jump. A global analysis (such as the
ones implemented in decompilers) could then supply the necessary context
to our handler analysis.
Unfortunately, Ghidra currently does not allow dynamic updates to the
language specification; thus, we leave the integration to future work.

\phead{Encryption register identification.} To correctly identify the context
required for decoding and analyzing instructions, we use a
heuristic approach. \toolname{} tracks a list of current context locations,
starting with the opcode of an instruction. If the context
depends on VM state not yet defined as context, we add this state
to the context. Additionally, any state involved in addressing
registers is also marked as context. Lastly, whenever we modify the context
specification, \toolname{} restarts the analysis with the new context specification.

\subsection{Tigress-Specific Implementation}\label{section:tigress}

Tigress supports multiple methods for
obfuscating against the identification of handlers and VM state. Attacking
obfuscation against VM state detection in an automated way is not the goal
of our research and has been covered in the literature~\cite{sok,vmdoctor2025,_rotalume}.
Therefore, we do not enable these additional obfuscations and
focus only on one dispatching method. In particular, we use the
\texttt{ifnest} dispatching method, which generates a dispatcher
loop with a series of nested if statements, one for each opcode.

Our preprocessing performs the following steps: First, we identify
the loop header address by finding the first conditional
branching instruction in the function code. Based on the loop header address,
we identify the location of state variables by inspecting the
initialization code before the loop header. Finally, we extract the
handler boundaries by analyzing the control flow after the loop header.
We implement this in 656 Python LOC using the \texttt{miasm} framework~\cite{miasm}.

\subsection{Code-Virtualizer-Specific Implementation}\label{section:codevirtualizer}

For the handler detection in Code Virtualizer, we implemented a
handler module for the table dispatch method. In particular, we analyze
the expression of symbolized, unconstrained jumps at the end of
the symbolic analysis. We perform pattern matching to determine whether
the jump target is loaded from the dispatcher table, and extract
the opcode formula.

If the jump target is not another handler, it may be a native transition, or obfuscated control-flow,
as depicted in \autoref{figure:jmp_obfuscation}. We use context concretization (c.f. \autoref{subsection:handler_analysis})
to address both cases. To mitigate path explosion, we also concretize the
context if the static branch count exceeds 50.

\begin{figure}
	\centering
	\caption{Jump obfuscation pattern used in Code Virtualizer, \textit{f} and \textit{g} represent
		arbitrary expressions based on the VM state and instruction operands. If \texttt{state} or \texttt{ops}
		is unknown, the target of the jump cannot be resolved.
       }
	\label{figure:jmp_obfuscation}
	\begin{lstlisting}[language=pseudoAsm,basicstyle=\fontfamily{SourceCodePro-TLF}\selectfont\scriptsize]
state[c] = jmptarget
state[f(state,ops)] = g(state,ops)
push state[c]
ret
	\end{lstlisting}
\end{figure}

For state identification, we implemented a script in 662 Python LoC 
based on the techniques discussed in prior work~\cite{vmdoctor2025,_rotalume}.
We generate a trace of up to 60 indirect jumps and take a register
snapshot at each. The VM state register always points into the binary and
remains constant across handlers, facilitating its identification.
Once the base VM state register is identified, we run another
trace and record all reads and writes to the VM state. The VPC is then
identified as the state register that always points into the binary
and is most often written to. The dispatch table and image base register
is identified by inspecting the VM prologue.

\subsection{SLEIGH Specification Generation}\label{subsection:sleigh_codegen}

In our prototype, we define the behavior and encoding of each VM instruction
in SLEIGH.  As discussed in \autoref{subsection:sleigh}, SLEIGH provides
processor specification engineers with multiple options to design
the specification such that the disassembler can efficiently perform
constant propagation of operand values. Currently, we do not
generate an optimized representation and
leave this to future work. Instead, we implement an additional
post-disassembler pass for constant propagation and dead branch removal
in Ghidra.

Although Ghidra's decompiler performs similar optimizations,
our approach provides two additional advantages. Firstly, optimized
\mbox{P-Code} helps reverse engineers better understand the effect of individual
instructions. Secondly, Ghidra's decompiler may interpret a
jump as a switch table when the corresponding indirect goto is resolved too late,
which frequently arises with expressions involving decryption. Our constant propagation
pass avoids this problem by optimizing the P-Code before it is handed to the decompiler.

\begin{figure}

	\begin{subfigure}{\columnwidth}
        \caption{Initial specification generated by \toolname{}.}
		\begin{lstlisting}[language=sleigh, numbers=none]
:Jcc opsize, cond, reloc32 is opcode=0xf; opsize; cond; reloc32 {
	if (opsize == 0) goto <opsize0>
	if (opsize == 1) goto <opsize1>
	if (opsize == 2) goto <opsize2>
	goto <opsize3>
	...
<opsize2>
	if (cond == 0) goto <cond0>
	if (cond == 1) goto <cond1>
	...
	goto <condX>
<cond0>
	if (OF != 1) goto <end>
	local _reloc:4 = reloc32;
	local target:8 = inst_start+6+sext(_reloc);
	goto [target];
   ...
<end>
   ...
}
		\end{lstlisting}
	\end{subfigure}

	\begin{subfigure}{\columnwidth}
		\caption{Specification after constant propagation and dead branch removal.}
		\begin{lstlisting}[language=sleigh, numbers=none]
# inst_start = 0x401023, opsize=2, cond=0, reloc32=0x20
:Jcc opsize, cond, reloc32 is ... {
	if (OF != 1) goto <end>
	goto 0x401049;
<end>
}
		\end{lstlisting}

	\end{subfigure}

        \caption{Simplified SLEIGH specification for a conditional jump instruction as generated by \toolname{}.}
	\label{figure:cjmp_example_us}
\end{figure}

In \autoref{figure:cjmp_example_us}, we depict how \toolname{}
would initially represent the conditional jump instruction presented in
\autoref{figure:cjmp_example}. For this example, we assume there is one
handler that computes the conditional jump for each \texttt{opsize}
and \texttt{cond}. As such, the generated code includes
sections for all possible conditions and opsize values.
Furthermore, since the jump destination
requires computations, we can only generate an indirect branch,
as denoted by the squared brackets.

While the initially generated code is large and complex, our post-disassembler
constant propagation simplifies the P-Code to just two instructions,
with no indirect branches.

Lastly, to specify disassembler \textit{context} updates, such as
encryption state, SLEIGH allows context modifications to be expressed in
disassembler expressions. However, it does not support updating the
context based on control-flow decisions in the P-Code (such as branch outcomes).
To address this limitation, our plugin includes a lightweight analyzer
that updates each possible successor address based on the
context modifications in the P-Code. Ghidra's existing processor plugins
also employ this approach.

\section{Evaluation}\label{section:evaluation}

To evaluate our prototype \toolname{}, we answer the following research questions:
\begin{itemize}[align=left,widest={\RQ{1}:},leftmargin=*]
	\setlength\itemsep{0pt}
	\setlength\parsep{3pt}
    \item[\RQ{1}:] Does \toolname{} produce semantically correct code?
    \item[\RQ{2}:] Is the decompiled code produced by \toolname{} similar to the original, and how does it compare to state-of-the-art VM-deobfuscators?
    \item[\RQ{3}:] What is the computational cost of applying \toolname{}?
\end{itemize}

\subsection{Experiment Setup}\label{section:experimental_method}

Our goal is to evaluate \toolname{} in a representative setting, demonstrating
applicability to a wide variety of programs and real-world payloads. For this
purpose, we obfuscate both a standard benchmark as well as algorithms extracted from
real malware. For obfuscators, we utilize the academic state-of-the-art obfuscator Tigress~\cite{tigress}
and the commercial obfuscator Code Virtualizer~\cite{codevirtualizer}.

We choose Code Virtualizer~\cite{codevirtualizer} as it uses the same virtualization technology as the Themida~\cite{themida} obfuscator~\cite{vmdoctor2025},
therefore, it is representative of both.
Moreover, current limitations of \toolname{}'s implementation (cf. ~\autoref{section:discussion_limitations}) -- in particular, state identification
and program context analysis -- prevent the automated analysis of VMProtect~\cite{vmprotect}.
Nevertheless, prior work has found that the handlers of Code Virtualizer are significantly
more complex~\cite{vmdoctor2025}. Therefore, we consider Code Virtualizer a sufficient representative
to evaluate \toolname{}'s capabilities in extracting the semantics of handlers produced by commercial obfuscators, given VM state information.

To gauge the readability of the code produced by \toolname{}, we qualitatively compare its output to the original source code,
the decompilation of unobfuscated code, and the output of the work by \mbox{Salwan et al.}~\cite{_tigress_deobf},
which is the only available state-of-the-art tool that produces simplified code.

We ran all experiments on a desktop machine with an Intel Xeon Silver 4110 CPU and 96 GB of DDR4 RAM.
We pinned each analysis to 4 CPU cores and allowed up to 32GB virtual address
space, running up to 3 analyses in parallel. 

\subsubsection{Program Sets}

We use five program sets as a basis to answer the different research questions.
In particular, we include program sets that contain complex control-flow, including input-dependent
loops and recursion.
Program sets A-D are based on the work of
Banescu et al.~\cite{banescu2016samples,banescu2017resilience} on benchmarking obfuscation approaches.
We slightly adjust them by removing any undefined behavior, since if the C semantics of the program are not well-defined,
source-based obfuscation, as well as compilation, changes the
original semantics, preventing us from validating our decompilation results.
Program set E is a reimplementation of six domain generation algorithms stemming from real-world malware.

\phead{Program Set A: Hash functions.}
This program set contains several different simple hash algorithms.
Due to the ease of validating hash functions, this set has also
been considered in several previous works~\cite{sok,vmdoctor2025}.

\phead{Program Set B: Small Programs.}
This program set contains several small single-function programs
that were specifically crafted to include a wide variety of code
characteristics. In particular, the following metrics were
varied: (1) size of the program input, (2) depth of nested control flow,
(3) number of if-statements, (5) number of loops,
(6) number of loops dependent on input size, and (7) number of loops dependent
on input values~\cite{banescu2016samples}.

\phead{Program Set C: Standard Algorithms.}
These include algorithms with more complex control flow, such as
input-dependent loops and recursive functions. Therefore, for most
programs, there exists an infinite number of paths.

\phead{Program Set D: Random Functions.}
Each program in this set consists of a single random function
that computes an output value based on a set of inputs. The control flow,
size, and type of input were varied. The original dataset contained
more than 5000 functions -- a size geared towards machine learning training~\cite{banescu2017resilience}.
Since several functions also included out-of-bounds reads due to the
number of input arguments read not matching the provided input, we randomly
selected and included 50 of these functions after manually correcting their behavior.

\phead{Program Set E: Domain Generation Algorithms.}
Lastly, to include real-world payloads, we reimplement DGAs based
on the Python reimplementation from Johann Bader's collection~\cite{bader_dga_github}.
Specifically, we chose three families (Zeus, Ramdo, Pushdo) featured in prior work
on taxonomy and measurement studies of DGAs~\cite{dga_taxonomy2016,dga_measurement2016}.
Further, we included the three most recent additions to this collection.
Command \& Control (C2) malware uses DGAs to compute the domain
of the C2 server. Consequently, an important goal of a malware analyst is
to extract the malware's DGA algorithm, enabling incident responders to filter
out future domains and prevent communication with the C2 server.

\begin{table}
	\centering
    \setlength{\tabcolsep}{3pt}
	\caption{Metrics of Program Sets}
	\label{figure:program-sets}
    \resizebox{\linewidth}{!}{%
	\begin{tabular}{c r r  r r r  r r r  r r r}
	\toprule
					 &            &             & \multicolumn{3}{c}{\#LOC} & \multicolumn{3}{c}{\#Tokens} & \multicolumn{3}{c}{CCN} \\
	Set				 &  \#Samples & \#Functions & min  &  max   &    mean     & min  &  max   &    mean        & min  &  max   &   mean   \\
	\midrule
		A         & 10        & 10          & 10   &  19    &   12.1     & 58   & 172    & 79.9          & 2    &  3     &   2.2   \\
		B         & 48        & 48          & 7    &  15    &   12.0     & 32   & 100    & 69.8          & 2    &  4     &   2.9   \\
		C         & 39        & 54          & 3    &  81    &   16.1     & 15   & 506    & 111.7         & 1    &  20    &   4.8   \\
		D         & 50        & 50          & 11   &  65    &   25.9     & 75   & 1081   & 309.4         & 2    &  6     &   3.7   \\
		E         & 6         & 26          & 3    &  50    &   17.2     & 27   & 357    & 143.0         & 1    &  9     &   3.8   \\
	\bottomrule
	\end{tabular}
    }
\end{table}

In \autoref{figure:program-sets}, we present complexity metrics of the program sets.
In particular, we depict the number of programs and functions,
minimum, maximum, and mean branch counts, as measured by the cyclomatic complexity number (CCN)~\cite{cyclomatic_complexity},
lines of code, and number of lexer tokens per function.

\subsubsection{Sample Sets}
To construct our obfuscated sample sets, we obfuscate each program set
using six configurations of Tigress v3.1\footnote{This is the latest 3.X version that successfully obfuscated all programs.}
and two configurations of Code Virtualizer v3.2.4.10.

We configure Tigress to use a stack-based, register-based, or mixed VM architecture.
For each architecture, we obfuscate using both the \emph{default}
configuration and the \mbox{\textit{SuperOperators}} configuration, which generates longer
and more complex bytecode instructions by merging instructions~\cite{tigress_website}.
For the concrete parameters, we followed the suggestions on the Tigress website~\cite{tigress_website}.
Lastly, we use a constant seed for reproducibility. We show the full commands in \apdxref{appendix:tigress_opts}.

For Code Virtualizer, we obfuscate the programs with \textit{Tiger Red} and \textit{Fish Red}.
\textit{Tiger Red} is Code Virtualizer's default configuration, which Zhang et al.~\cite{vmdoctor2025}
identified as using a register-based RISC architecture. To also include a CISC variant,
we additionally apply the \textit{Fish Red} configuration, which contains
far longer and more complex instructions~\cite{vmdoctor2025}. We stick to the \textit{Red} sub-configuration
as it is the default, and prior work found that the color only trivially changes obfuscation~\cite{vmdoctor2025}.
To obfuscate our programs, we insert Code Virtualizer's obfuscation macros at the beginning and end
of each function. To ensure that the compiler actually places the basic blocks containing
the macros at the beginning and end, we turn off reordering optimizations.
We include the complete command line in \apdxref{appendix:codevirtualizer_sample_opts}.

In \autoref{figure:obfuscation_options_complexity}, we present the minimum, maximum, and mean counts
of handlers and virtual instructions, as well as handler complexity, measured by the number of instructions and cyclomatic 
complexity, across all sample sets for the different Tigress and Code Virtualizer configuration options.

\begin{table*}[h]
	\centering
	\caption{Metrics per obfuscation options over all samples}
	\label{figure:obfuscation_options_complexity}
    \resizebox{\linewidth}{!}{%
	\begin{tabular}{l l r r r  r r r  r r r  r r r }
		\toprule
		Obfuscator & Configuration & \multicolumn{6}{c}{VM-Complexity} & \multicolumn{6}{c}{Handler Complexity}  \\
		                          && \multicolumn{3}{c}{\#Handlers} & \multicolumn{3}{c}{\#VMInstructions} & \multicolumn{3}{c}{\#NativeInstructions} & \multicolumn{3}{c}{CCN} \\
		                          && min     &  max     &    mean     & min  &  max  &  mean                   & min      &    max   &    mean              & min  &  max   &   mean   \\
		\midrule

Tigress & Stack-Default & 7 & 42 & 25.0 & 20 & 938 & 193.4& 5 & 53 & 12.6 & 1 & 4 & 1.0 \\
Tigress & Stack-SuperOperators & 7 & 91 & 23.7 & 14 & 224 & 63.9& 5 & 101 & 50.0 & 1 & 4 & 1.0 \\
Tigress & Register-Default & 7 & 42 & 25.0 & 20 & 938 & 193.4& 5 & 53 & 16.9 & 1 & 3 & 1.0 \\
Tigress & Register-SuperOperators & 7 & 104 & 28.0 & 14 & 619 & 79.9& 5 & 154 & 62.6 & 1 & 4 & 1.0 \\
Tigress & Mixed-Default & 9 & 49 & 29.9 & 29 & 1436 & 276.7& 5 & 38 & 15.6 & 1 & 4 & 1.0 \\
Tigress & Mixed-SuperOperators & 8 & 135 & 37.3 & 16 & 886 & 111.3& 5 & 119 & 53.1 & 1 & 4 & 1.0 \\
CodeVirtualizer & Tiger-Red & 17 & 136 & 48.9 & 43 & 2787 & 311.4& 28 & 4315 & 349.9 & 1 & 40 & 4.2 \\
CodeVirtualizer & Fish-Red & 16 & 94 & 42.2 & 45 & 2821 & 302.6& 88 & 7747 & 1382.3 & 1 & 135 & 23.9 \\

		\bottomrule
	\end{tabular}
}
\end{table*}

\subsection{\RQ{1}: Semantic Correctness}

To validate the semantic correctness, we
apply \toolname{} to deobfuscate a sample and then recompile
the decompilation output.
To allow for recompilation, we define Ghidra's non-standard types
and perform regex replacements to remove common decompilation artifacts.
For global variables, we reuse the definitions from
the original source file.

We validate the functional correctness using randomized input-output testing.
To generate meaningful inputs, we manually define the number and
expected types of program arguments. For example, several programs use \texttt{strtol}
to convert an argument to a number. If we do not pass a valid number, these functions
return an error and abort the program without reaching any deeper components.
Additionally, we manually specify inputs to include all error cases. Lastly, we collect
and inspect the branch coverage of our testing method and validate that we cover
all feasible branches in the original binary. In total, we observed that 
1000 inputs with this method suffice to cover all branches in the sample programs.

We present a summary of the decompilation results in \autoref{table:eval_summary}.
\toolname{} was able to decompile all 918 Tigress samples and produce semantically
correct code. For Code Virtualizer Tiger, 152 of 153 samples could be correctly
decompiled. The remaining sample is \texttt{numtowords}, featuring a jump table.
Since Code Virtualizer does not yet support jump tables, VM obfuscation is applied before
and after the jump table. When analyzing the semantics of the native transition handler,
the target of the jump table cannot be resolved without additional program context since the jump table
location and index are computed inside the virtualized code. For the same sample in Tigress,
the jump table is included in the virtual bytecode, enabling Ghidra to recover
it directly.

Lastly, for Code Virtualizer Fish, 149 of 153 samples could be correctly decompiled.
Two samples could not be analyzed because the analysis ran out of memory.
Additionally, \texttt{numtowords} failed for the same reason as mentioned above.
For the remaining sample, our context identification was too imprecise, leading to missed handlers.

\begin{table}[h]
    \resizebox{\linewidth}{!}{%
	\begin{tabular}{l  c c c  l l  c  c c c  c c c }
		\toprule
		\textbf{Set} & \multicolumn{3}{c}{\textbf{Source}} & \textbf{Obfuscator} & \textbf{Configuration} & \textbf{Success} & \multicolumn{3}{c}{\textbf{Initial Decompilation}} & \multicolumn{3}{c}{\textbf{Final Decompilation}} \\
				& LOC & Tokens & CCN                  &	         &                 &             & LOC & Tokens & CCN & LOC & Tokens & CCN \\

\midrule \midrule
\multirow{9}*{A} &  \multirow{9}*{12.1} & \multirow{9}*{79.9} & \multirow{9}*{2.2} & \multirow{1}*{Original} & -- & -- / -- & -- & -- & -- &17.7 & 84.8 & 3.2\\

	\cline{5-13}
 &  & & & \multirow{6}*{Tigress} & StackDefault & 10 / 10 & 20.8 & 131.7 & 3.0 & 17.7 & 84.1 & 3.2\\
 &  & & & & StackSuperOp & 10 / 10 & 20.8 & 131.7 & 3.0 & 17.7 & 84.1 & 3.2\\
 &  & & & & RegDefault & 10 / 10 & 18.4 & 126.9 & 2.2 & 17.7 & 84.1 & 3.2\\
 &  & & & & RegSuperOp & 10 / 10 & 18.4 & 126.9 & 2.2 & 17.7 & 84.1 & 3.2\\
 &  & & & & MixedDefault & 10 / 10 & 20.8 & 131.7 & 3.0 & 17.7 & 84.1 & 3.2\\
 &  & & & & MixedSuperOp & 10 / 10 & 20.5 & 131.1 & 2.9 & 17.7 & 84.1 & 3.2\\

	\cline{5-13}
 &  & & & \multirow{2}*{Code Virt.} & Tiger-Red & 10 / 10 & 20.3 & 101.3 & 3.2 & 17.4 & 85.4 & 3.2\\
 &  & & & & Fish-Red & 10 / 10 & 19.4 & 94.1 & 3.2 & 17.5 & 86.1 & 3.2\\
\midrule \midrule
\multirow{9}*{B} &  \multirow{9}*{12.0} & \multirow{9}*{69.8} & \multirow{9}*{2.9} & \multirow{1}*{Original} & -- & -- / -- & -- & -- & -- &19.2 & 106.7 & 3.9\\

	\cline{5-13}
 &  & & & \multirow{6}*{Tigress} & StackDefault & 48 / 48 & 20.2 & 131.1 & 3.5 & 20.9 & 125.1 & 4.0\\
 &  & & & & StackSuperOp & 48 / 48 & 20.1 & 130.8 & 3.5 & 20.8 & 124.8 & 4.0\\
 &  & & & & RegDefault & 48 / 48 & 18.4 & 127.7 & 2.9 & 21.1 & 125.6 & 4.0\\
 &  & & & & RegSuperOp & 48 / 48 & 18.4 & 127.7 & 2.9 & 21.1 & 125.6 & 4.0\\
 &  & & & & MixedDefault & 48 / 48 & 20.1 & 130.8 & 3.5 & 20.9 & 125.1 & 4.0\\
 &  & & & & MixedSuperOp & 48 / 48 & 20.0 & 131.0 & 3.5 & 20.8 & 124.8 & 4.0\\

	\cline{5-13}
 &  & & & \multirow{2}*{Code Virt.} & Tiger-Red & 48 / 48 & 39.0 & 282.4 & 5.6 & 32.7 & 200.7 & 5.0\\
 &  & & & & Fish-Red & 47 / 48 & 38.4 & 298.9 & 5.7 & 35.8 & 227.3 & 4.8\\
\midrule \midrule
\multirow{9}*{C} &  \multirow{9}*{16.1} & \multirow{9}*{111.7} & \multirow{9}*{4.8} & \multirow{1}*{Original} & -- & -- / -- & -- & -- & -- &31.6 & 181.1 & 5.6\\

	\cline{5-13}
 &  & & & \multirow{6}*{Tigress} & StackDefault & 39 / 39 & 36.1 & 277.7 & 5.4 & 35.1 & 229.6 & 5.7\\
 &  & & & & StackSuperOp & 39 / 39 & 35.6 & 273.6 & 5.3 & 34.9 & 228.0 & 5.6\\
 &  & & & & RegDefault & 39 / 39 & 33.2 & 261.2 & 4.9 & 34.7 & 226.2 & 5.6\\
 &  & & & & RegSuperOp & 39 / 39 & 33.2 & 261.2 & 4.9 & 34.7 & 226.1 & 5.6\\
 &  & & & & MixedDefault & 39 / 39 & 36.3 & 279.0 & 5.3 & 35.1 & 230.1 & 5.5\\
 &  & & & & MixedSuperOp & 39 / 39 & 35.7 & 273.9 & 5.3 & 34.7 & 225.4 & 5.6\\

	\cline{5-13}
 &  & & & \multirow{2}*{Code Virt.} & Tiger-Red & 38 / 39 & 50.9 & 371.6 & 9.0 & 45.1 & 299.8 & 6.5\\
 &  & & & & Fish-Red & 36 / 39 & 50.6 & 371.3 & 8.9 & 40.8 & 274.7 & 6.4\\
\midrule \midrule
\multirow{9}*{D} &  \multirow{9}*{25.9} & \multirow{9}*{309.4} & \multirow{9}*{3.7} & \multirow{1}*{Original} & -- & -- / -- & -- & -- & -- &13.0 & 92.9 & 2.3\\

	\cline{5-13}
 &  & & & \multirow{6}*{Tigress} & StackDefault & 50 / 50 & 22.4 & 211.9 & 2.9 & 13.9 & 106.9 & 2.3\\
 &  & & & & StackSuperOp & 50 / 50 & 21.3 & 207.5 & 2.9 & 14.3 & 109.0 & 2.3\\
 &  & & & & RegDefault & 50 / 50 & 20.8 & 206.4 & 2.6 & 13.9 & 107.6 & 2.3\\
 &  & & & & RegSuperOp & 50 / 50 & 20.8 & 206.4 & 2.6 & 13.9 & 107.8 & 2.3\\
 &  & & & & MixedDefault & 50 / 50 & 22.6 & 219.5 & 2.9 & 14.5 & 112.1 & 2.3\\
 &  & & & & MixedSuperOp & 50 / 50 & 21.4 & 208.1 & 2.9 & 13.9 & 106.1 & 2.3\\

	\cline{5-13}
 &  & & & \multirow{2}*{Code Virt.} & Tiger-Red & 50 / 50 & 17.4 & 121.7 & 2.3 & 14.5 & 104.1 & 2.3\\
 &  & & & & Fish-Red & 50 / 50 & 18.1 & 130.2 & 2.3 & 15.0 & 112.9 & 2.3\\
\midrule \midrule
\multirow{9}*{E} &  \multirow{9}*{17.2} & \multirow{9}*{143.0} & \multirow{9}*{3.8} & \multirow{1}*{Original} & -- & -- / -- & -- & -- & -- &34.0 & 214.7 & 4.8\\

	\cline{5-13}
 &  & & & \multirow{6}*{Tigress} & StackDefault & 6 / 6 & 43.8 & 338.3 & 4.3 & 42.9 & 254.7 & 4.9\\
 &  & & & & StackSuperOp & 6 / 6 & 49.7 & 372.2 & 4.4 & 48.8 & 279.3 & 4.9\\
 &  & & & & RegDefault & 6 / 6 & 47.4 & 358.5 & 4.0 & 50.0 & 286.6 & 4.9\\
 &  & & & & RegSuperOp & 6 / 6 & 47.4 & 358.5 & 4.0 & 50.0 & 286.6 & 4.9\\
 &  & & & & MixedDefault & 6 / 6 & 46.3 & 349.0 & 4.3 & 45.0 & 263.7 & 5.0\\
 &  & & & & MixedSuperOp & 6 / 6 & 49.4 & 366.9 & 4.4 & 47.7 & 275.6 & 4.7\\

	\cline{5-13}
 &  & & & \multirow{2}*{Code Virt.} & Tiger-Red & 6 / 6 & 46.9 & 329.9 & 6.4 & 42.7 & 267.5 & 4.8\\
 &  & & & & Fish-Red & 6 / 6 & 47.0 & 332.0 & 6.3 & 43.2 & 270.6 & 4.8\\

		\bottomrule
	\end{tabular}%
    }
	\caption{Summary of \toolname{}'s decompilation results. Original denotes unobfuscated code. Initial decompilation denotes
	the initial output from \toolname{}. Final decompilation denotes the output after compilation to x64 and decompilation with Ghidra.
	LOC, CCN and token is per obfuscated function.}
	\label{table:eval_summary}
\end{table}

\subsection{\RQ{2}: Code Similarity}\label{subsection:code_similarity}

In this section, we first manually inspect and qualitatively compare
the decompilation output of \toolname{} with the original source code and the output from the
DSE approach by Salwan et al.~\cite{_tigress_deobf} on the Tigress stack
VM. Salwan et al.'s tool is the only open-source state-of-the-art approach
that generates a deobfuscated binary.
However, their method does not work for programs with input-dependent
loops or recursive function calls, i.e., all program sets A-E. Therefore, we manually convert input-dependent loops
to constant-bounded versions of length 8 for program set A (i.e., hash functions). Additionally, the input program must
use \texttt{strtoul} for input and \texttt{printf} for output.
This closely mimics their own experimental evaluation~\cite{_tigress_deobf}.
For the adjusted program set A, the tool worked for 8 out of 10 hash functions.
Afterward, we compare complexity metrics for
the output of \toolname{} and Ghidra's x64 decompilation of unobfuscated code.

We depict the different deobfuscated outputs on the DEKHash sample obfuscated with
\emph{StackSuperOperator} in \autoref{figure:rq3_dekhash}. Specifically, we
compare the original C code, the unobfuscated x64 decompilation output, the output of
Salwan et al.'s~\cite{_tigress_deobf} approach, and \toolname{}'s decompilation.
In \autoref{subfigure:salwan}, since code is generated based on the direct relation
between program input and output bytes, we observe a loop-unrolled expression in the deobfuscated code.
The operation in each loop iteration
is still visible in the output. In contrast, \toolname{} (\autoref{subfigure:dekhash})
accurately recovers the control flow of
the original code. We can observe that the field \texttt{stk\_1c}
corresponds to the \texttt{hash} variable, \texttt{stk\_20} to
\texttt{i}, and \texttt{stk\_00} to \texttt{str}. The for loop has been converted to an equivalent
while loop. In comparison, the x64 decompilation more accurately recovers an actual for loop.
Overall, we consider the code recovered by \toolname{} very similar to the original.

\begin{figure*}[htb]
	\centering
    \caption{Comparison of original code, unobfuscated decompilation by Ghidra, decompiled code after deobfuscation by Salwan et al.~\cite{_tigress_deobf} and the output of \toolname{} after deobfuscation on the example of the DEKHash SuperOperator configuration. Variable names renamed for brevity.}
	\label{figure:rq3_dekhash}
	\begin{subfigure}{0.4\textwidth}
		\caption{Original DEKHash C code}
		\begin{lstlisting}[language=reC,basicstyle=\fontfamily{SourceCodePro-TLF}\selectfont\tiny]
unsigned int DEKHash(unsigned char* str, unsigned int len) {
   unsigned int hash = len;
   unsigned int i    = 0;
   for(i = 0; i < len; str++, i++) {
      hash = ((hash << 5) ^ (hash >> 0x1b)) ^ (*str);
   }
   return hash;
}
		\end{lstlisting}
	\end{subfigure}
	\hspace{6em}
	\begin{subfigure}{0.35\textwidth}
		\caption{Ghidra decompilation of unobfuscated code.}
		\begin{lstlisting}[language=reC,basicstyle=\fontfamily{SourceCodePro-TLF}\selectfont\tiny]
uint DEKHash(byte *param_1,uint param_2) {
  byte *l_20;
  uint l_10;
  uint l_c;
  l_20 = param_1;
  l_c = param_2;
  for (l_10 = 0; l_10 < param_2; l_10 = l_10 + 1) {
    l_c = (uint)*l_20 ^ (l_c << 5 | l_c >> 0x1b);
    l_20 = l_20 + 1;
  }
  return l_c;
}
		\end{lstlisting}
	\end{subfigure}

	\begin{subfigure}{0.4\textwidth}
		\caption{Deobfuscation with \mbox{Salwan et al.~\cite{_tigress_deobf}'s} method.\newline Decompiled with Ghidra.}
		\label{subfigure:salwan}
		\begin{lstlisting}[language=reC,basicstyle=\fontfamily{SourceCodePro-TLF}\selectfont\tiny]
uint DEKHash(undefined8 param_1) {
  uint uVar1;
  uint uVar2;
  uVar2 = ((uint)param_1 & 0xff00) << 0xc ^ (uint)param_1 << 0x19;
  uVar1 = ((uint)(param_1 >> 0x28) & 0xff ^
          ((uint)(param_1 >> 7) & 1) +
          ((uint)(param_1 >> 0x1b) & 0x1fe0 ^ uVar2 ^
          ((uint)(param_1 >> 0x18) & 0xff ^ (uint)(param_1 >> 0xb)
            & 0x1fe0) << 10) + 2) << 5;
  return (((ushort)(param_1 >> 0x30) & 0xff ^
         (uVar1 | uVar2 >> 0x1b)) << 5 | uVar1 >> 0x1b) ^
         (uint)(byte)(param_1 >> 0x38);
}
		\end{lstlisting}
	\end{subfigure}
	\hspace{6em}
	\begin{subfigure}{0.35\textwidth}
        \caption{\toolname{}'s decompilation output.}
		\label{subfigure:dekhash}
		\begin{lstlisting}[language=reC,basicstyle=\fontfamily{SourceCodePro-TLF}\selectfont\tiny]
uint DEKHash(byte *param_1,uint param_2)
{
  byte *stk_00;
  uint stk_1c;
  uint stk_20;
  stk_20 = 0;
  stk_00 = param_1;
  stk_1c = param_2;
  while( true ) {
	 if (param_2 <= stk_20) break;
	 stk_1c = stk_1c >> 0x1b ^ stk_1c << 5 ^ (uint)*stk_00;
	 stk_00 = stk_00 + 1;
	 stk_20 = stk_20 + 1;
  }
  return stk_1c;
}
		\end{lstlisting}
	\end{subfigure}

\end{figure*}

To illustrate a slightly more complex example, we depict the output
of \toolname{} on the recursive binary search implementation obfuscated
with Tigress MixedSuperOperators in
\autoref{figure:rq3_binarysearchrec}. This sample includes several
more complex features, such as arrays and recursive calls.
We observe that Ghidra fails to reconstruct the array access \texttt{a[mid]}
and instead produces a dereference operation on a pointer computation. While this
is semantically equivalent, it impacts readability for humans.
We observed this behavior also when decompiling the unobfuscated x64
binary. Such type reconstruction represents a
generally hard challenge for decompilers~\cite{2020_liu_decompilers} and 
is a general Ghidra shortcoming, not specific to our design.
On the same sample obfuscated with Code Virtualizer Fish, we observe
several additional artifacts in the output that may impact
readability. We depict an example in \autoref{appendix:rq3_binarysearchrec_cv} of the Appendix.

\begin{figure}[htb]
	\centering
	\caption{Comparison of original code, and output of \toolname{} on \texttt{binarysearchrec} obfuscated with Tigress Mixed-SuperOperator.}
	\label{figure:rq3_binarysearchrec}

	\tabskip=0pt
	\begin{subfigure}{0.4\textwidth}
		\caption{Original recursive binary search C Code}
		\begin{lstlisting}[language=reC,basicstyle=\fontfamily{SourceCodePro-TLF}\selectfont\tiny]
int binary(int a[],int n,int m,int l,int u) {
     int mid,c=0;
     if(l < u){
          mid = (l+u)/2;
          if(m==a[mid]){
              c = 1;
          }
          else if(m < a[mid]){
              return binary(a,n,m,l,mid-1);
          }
          else
              return binary(a,n,m,mid+1,u);
     }
     return c;
}
		\end{lstlisting}
	\end{subfigure}
	\begin{subfigure}{0.4\textwidth}
		\caption{\toolname{}'s output after deobfuscation.}
		\label{subfigure:rq3_binsearch_out}
		\begin{lstlisting}[language=reC,basicstyle=\fontfamily{SourceCodePro-TLF}\selectfont\tiny]
undefined4 binary(long long param_1, undefined4 param_2, int param_3, int param_4, int param_5)
{
  int iVar1;
  undefined4 uVar2;
  struct __attribute__((packed)) {
    undefined4 uStack_10040;
    undefined1 uStackPad1003c[65596];
  } stack;

  stack.uStack_10040 = 0;
  if (param_4 < param_5) {
    iVar1 = (param_5 + param_4) / 2;
    if (param_3 != *(int *)((longlong)iVar1 * 4 + param_1)) {
      if (*(int *)((longlong)iVar1 * 4 + param_1) <= param_3) {
        uVar2 = binary(param_1,param_2,param_3,iVar1 + 1,param_5,0);
        return uVar2;
      }
      uVar2 = binary(param_1,param_2,param_3,param_4,iVar1 + -1,0);
      return uVar2;
    }
    stack.uStack_10040 = 1;
  }
  return stack.uStack_10040;
}
		\end{lstlisting}
	\end{subfigure}
\end{figure}

In summary, we observe the following artifacts in our output: First,
Ghidra always inferred a return value and type, even if the original
function did not have a return value. Next, the recovered types,
especially for more complex types such as pointers and arrays,
are inaccurate and cannot be inferred; however, this does
not impact the semantics (ignoring C's aliasing rules).
Additionally, for Code Virtualizer, we observe that Ghidra
fails to simplify more expressions, such as comparisons and
signed operations. This occurs because Code Virtualizer's obfuscation
leads to slightly modified expressions that are missed by
Ghidra's existing simplification rules.

To investigate the extent to which decompilation artifacts stem from
Ghidra or \toolname{}, we compare the complexity metrics (LOC, Lexer tokens, CCN)
for the decompilation of the deobfuscated and original x64 code.
Additionally, we assess whether compiler optimizations can
remove artifacts that Ghidra cannot remove. Therefore,
we also compare code complexity after recompilation
to x64 and following decompilation with Ghidra. We list
the used compilation flags in \apdxref{appendix:redec_opts} and
present the results in \autoref{table:eval_summary}.

In all program sets except D, the resulting decompiled code is longer and
more verbose than the original source code, which we explain with the 
previously shown artifacts. For program set D, we find that the decompiled code is, on average,
less complex than the original source code. Upon inspection, we find that the randomized programs contain many
lines of dead code and redundant statements, which were eliminated by the (de-)compiler passes.
For the simple hash program set, we observe that the output of our final output and the decompilation of the original
are identical in nearly all cases. For the remaining, more complex program sets,
the final outputs are slightly more complex in LOC and number of lexer tokens compared to the decompilation
of the original. Overall, recompilation followed by Ghidra's x64 decompilation
reduces the number of lines and lexer tokens. For the Tigress samples, the branch count increases with optimized
recompilation (i.e., increased CCN), whereas it decreases for the Code Virtualizer samples.

\subsection{\RQ{3}: Computational Cost}

In this section, we investigate the computational cost of handler analysis and decompilation
for individual obfuscated samples. In particular, we are interested in how the
cost scales with the complexity of the VM and the obfuscated sample. For all analyzed samples, we depict their handler analysis runtime and their VM complexity, as indicated by the sum of all handler instructions, in \autoref{figure:perf_scatter_plot_handler_analysis}. Similarly, we depict the decompilation time in \autoref{figure:perf_scatter_plot_decompile}.

\input{./perf_scatter_plot.tex}

\input{./perf_scatter_plot_decompile.tex}

\phead{Handler Analysis.} For all Tigress and Code Virtualizer samples, the handler analysis time scales roughly linearly with the number of handler instructions. Compared to the Tigress samples, Code Virtualizer requires one magnitude more runtime per handler instruction. This relates to the higher branch count (as measured by CCN), which causes the analysis to iterate over the handler instructions in all handler paths several times. In particular, the encryption logic inside Code Virtualizer adds many additional branches. This effect is further enhanced for Code Virtualizer-Fish, which follows a CISC architecture and, therefore, includes more complex handlers and more encryption key updates. As a consequence of these additional branches, symbolic execution re-examines the handler's core semantics multiple times, blowing up the execution time. This is a well-known challenge that symbolic execution approaches struggle with. We discuss implications in \autoref{section:discussion_limitations}. %

\phead{Decompilation.} For decompilation, no clear relationship between decompilation time and VM complexity is visible. In particular, the majority of decompilation runs require between 40 and 60 seconds. Some samples obfuscated with Code Virtualizer-Fish require more time to decompile, the largest one taking 170.8 seconds. For most samples, the majority of the time is spent on Ghidra's \textit{startup}, which includes loading all generated plugins. Overall, \toolname{}'s performance is practical for deeper inspection of VM-obfuscated binaries, even for modern, commercial VM obfuscators.

\section{Discussion}\label{section:discussion}

In this section, we discuss potential future work, limitations, and
countermeasures to \toolname{}.

\subsection{Integration With Other Frameworks}

\toolname{}'s generated SLEIGH processor specifications
allow reverse engineers to directly leverage all analysis frameworks
that support SLEIGH. This includes the
popular DSE engine angr~\cite{angr} and binary analysis
frameworks such as radare2~\cite{radare2} and BAP~\cite{bap}.
These frameworks already implement several other analysis and deobfuscation
techniques, which could enable
defeating additional obfuscations in the VM bytecode. It is up to
future work to evaluate these combinations and, specifically, whether
further analysis of the decompiled code or \mbox{P-Code} proves
more beneficial.

\subsection{Limitations}\label{section:discussion_limitations}

\phead{Path Explosion.}
The complexity of current commercial handlers can already lead to
path explosion. While our context concretization approach allows us
to analyze the vast majority of handlers, it increases the analysis time and memory
consumption. This is
an inherent limitation of DSE. Prior work has suggested merging
states~\cite{statemerging2012} or combining DSE with static symbolic execution~\cite{veritesting}
to mitigate this issue. We leave integration with more efficient approaches
to future work.

\phead{Program-Context-based Specialization.}
Our current prototype of \toolname{} only considers instruction-local
context for specialization of native transitions, i.e., encryption
registers, VPC value, and operands. However, native transitions
may require more program context to deduce the target of a native
transition or indirect jumps in the native code. We have observed
this in the \texttt{numtowords} sample and, generally, in the
native transitions of VMProtect. In particular, VMProtect pushes the
target of the native transition in one VM instruction and uses it in
another, requiring an analysis to include the stack context.
However, this is not a fundamental limitation since existing reverse-engineering
frameworks such as Ghidra already perform sufficient analysis to infer that context.
We leave better integration to future work.

\phead{State identification effort.}
Adding support for additional VM obfuscators currently requires
additional efforts to identify and model relevant VM state.
While prior work discussed automated methods~\cite{vmdoctor2025,_rotalume,_pushan}, no
implementations are available. Integration with generic
methods is part of future work.

\phead{Arithmetic Obfuscations.}
\toolname{} applies symbolic reasoning at several stages of the
analysis. For example, the memory model and graph optimizations
assume it is possible to identify constant offsets to base
pointers. Arithmetic obfuscations can hinder this analysis.
While \toolname{} includes simplifications based on simple arithmetic
properties (such as $a-a=0$), more complex arithmetic obfuscations,
in particular Mixed Boolean-Arithmetic (MBA)~\cite{mba}, can hinder
this analysis. Therefore, \toolname{} depends on an expression
simplifier. However, we leave integration with MBA-deobfuscation
approaches~\cite{_syntia,xyntia,xsmir,mbablast,arybo} to future work.

\phead{General Decompiler Limitations.}
As \toolname{} leverages the capabilities of decompilers, it also inherits
their limitations. Lui et al.~\cite{2020_liu_decompilers} have
analyzed the capabilities and limitations of modern decompilers.
Most notably, decompilers still struggle in the classical
reverse-engineering problem of type recovery, including function
signature detection. Our evaluation confirms these results for
Ghidra. Academia proposes techniques to
address common issues such as subtyping, recursive types, and
polymorphism~\cite{2016_TIE_caballero,2016_retypd_noonan,2024_binsub_smith};
however, state-of-the-art decompilers have yet to adopt them.
Another common problem in decompilers is limited readability due to
over-optimization~\cite{2020_liu_decompilers}. Nevertheless,
as \toolname{} enables leverage the power of and improvements in
decompilers, its results will also improve with
advances in academia and industry.

\phead{Decompiler integration.}
VM obfuscation is usually applied only to selected functions
because it incurs high overhead. Therefore, if we consider each
randomized VM its own CPU architecture, the binary will contain
code for several of them that interact with each other. However, Ghidra currently
does not directly support a binary with multiple instruction sets~\cite{ghidra_multiple_procs}.
As a consequence, the VM code cannot be analyzed
alongside the native code. We leave better integration with the Ghidra
framework up to future work.

\phead{Partially Virtualized Code.}
We investigated VM obfuscation applied to whole functions. In some cases,
however, malware virtualizes only one or a few lines at a time.
While \toolname{} would allow for the decompilation of the snippet,
its semantics cannot be directly analyzed in the context of the embedding
function. While this can be resolved with better decompiler integration,
the approach might not be optimal. Specifically, prior work has shown
that due to their simplicity, such snippets can be effectively
analyzed with dynamic symbolic execution alone~\cite{_vmhunt}, and,
further, black-box deobfuscation approaches~\cite{_syntia,xyntia,xsmir}
show promising results that may be applied for deobfuscating smaller virtualized code.

\subsection{Potential Countermeasures}\label{subsection:countermeasures}

Potential countermeasures can target multiple steps of \toolname{}. Firstly,
our approach requires identifying the VM state, bytecode opcodes,
and their corresponding handlers. The VM state location may be
additionally obfuscated to prevent identification. In particular, regularly shuffling
the VM state's location may impede the currently proposed techniques~\cite{vmdoctor2025,_rotalume}.
While we have observed that VMProtect already implements this countermeasure,
PUSHAN~\cite{_pushan} demonstrated that it is feasible to re-identify the location
of VM state, in particular the VPC.

Secondly, the use of DSE in handler analysis is subject to further obfuscation 
specifically targeted against it. While a multitude of techniques
exist~\cite{ollivier2019kill,banescu2016samples,schloegel2022loki,collbergprobabilistic},
systematizations~\cite{ollivier2019obfuscation} were unable to make clear
statements on whether they fundamentally thwart DSE analysis.
Since bytecode handlers are substantially smaller than whole programs, we argue that detecting
and mitigating anti-DSE techniques is simpler at the handler level
than during whole-program analysis.

Thirdly, malware authors may specifically attack weaknesses of decompilers. In particular,
they can produce constructs that are difficult for decompilers to handle~\cite{dramko2024taxonomy}
or trigger implementation-specific bugs~\cite{2020_liu_decompilers}. Additionally, attackers may apply obfuscation methods to the 
underlying program logic, rendering the decompilation output difficult to understand.

\section{Related Work}\label{section:related_work}

\phead{Reverse-Engineering Virtual Machine Intrinsics.}
Rolles~\cite{_rolles} manually reverse-engineered VMProtect and
all its handlers, defined semantics in a custom IR, and used regular expressions
to identify corresponding handlers in other samples.
However, modern diversification techniques render this manual process
infeasible. \toolname{} also maps
handlers to IR (\mbox{P-Code}), but automates handler analysis to
generate processor specifications, leveraging
existing decompilers and reverse-engineering frameworks.

Sharif et al.~\cite{_rotalume} presented
generic automated methods to locate VM state and
handler boundaries, as well as to construct a control-flow graph. However,
the robustness of the underlying techniques to modern
diversification techniques is unclear~\cite{vmdoctor2025}. Zhang et al.~\cite{vmdoctor2025} studied
modern VM diversification techniques and
developed an automated tool to identify handler boundaries and
diversification techniques. Since \toolname{} relies on the location
of VM state and handler boundaries, it may be combined with these methods.

Cheng et al.~\cite{_dynopvm} propose a frequency analysis to map virtual
instructions to native ones. However, this attack does not work on
complex mappings or virtual instructions whose semantics are incomplete without operands.
\mbox{Shijia et al.~\cite{li2022chosen}} propose a chosen-instruction attack
to extract reusable information, such as the context-switch scheme and
dispatcher structure, to aid deobfuscation. This information could
be combined with \toolname{} in the preprocessing and handler detection steps.

\phead{Black-Box Deobfuscation.}
Other works~\cite{_syntia,xyntia,xsmir,david2020qsynth} apply black-box program synthesis
to simplify obfuscated expressions, in particular MBA expressions.
Since black-box deobfuscation relies only on input-output pairs, it
is resistant to most obfuscation schemes, including VM-based obfuscation.
However, current approaches can only be applied to simplify \textit{expressions}
and do not consider memory or control-flow semantics.
Two approaches~\cite{_syntia,xyntia} were successfully applied
to simplify expressions in VM handlers and to identify arithmetic
handlers. Overall, the handler analysis of \toolname{} could be
enhanced with black-box deobfuscation to resist arithmetic obfuscation.

\section{Conclusion}\label{section:conclusion}

We presented \toolname{}, a novel methodology for
deobfuscating virtualized code by leveraging existing decompilers.
To this end, we automate the generation of VM processor specifications
by using dynamic symbolic execution to extract instruction
encoding and semantics from VM bytecode handlers.
To evaluate our prototype of \toolname{}, built on top of the SLEIGH processor
specification language and Ghidra, we obfuscate \numprograms synthetic programs
and \numdgas DGAs extracted from real-world malware using
both academic and commercial VM obfuscators.
We find that \toolname{} produces semantically correct
and high-quality decompilation in nearly all cases.
In particular, \toolname{} is the first approach demonstrating
recovery of complete program semantics for VM-obfuscated
programs containing complex control flow such as input-dependent
loops.

\section*{Ethical Considerations}

This section details our ethical considerations for the research
performed in respect to \textit{respect for persons, beneficience, justice
and respect for law and interests}.
Relevant stakeholders we considered are the software consumers,
security practitioners, developers of obfuscation software and
software copyright holders.

Our research does not involve human participants or personal data.
Consequently, there is no potential for harm to individuals, and \textit{respect for persons} is inherently maintained.

Regarding \textit{beneficience}, advancing research into reverse engineering
benefits software consumers and security practitioners by aiding in
identifying malicious code, mitigating attacks early on and, therefore,
protecting software consumers.
We acknowledge the research may lead to harm to companies providing
obfuscation services and may be used for copyright infringement.
Overall, we believe benefits of being able to analyze and identify
malicious code outweighs the potential harm of mentioned stakeholders
and our work provides a net positive benefit.

Concerning \textit{justice}, our research did not include or
exclude persons or groups based on any attributes of a given
person or group.

Regarding respect for law and interests, we considered licensing
restrictions of the mentioned Tigress, Code Virtualizer and VMProtect
softwares. The licenses do not prohibit the reverse engineering of protected
software. Further, we did not perform any reverse engineering of
the protection software itself nor violated any other licensing
agreements.

\section*{Acknowledgements}

We would like to thank the anonymous reviewers for their
time and insightful feedback. We further thank Fabian Franzen
and Maximilian von Tschirschnitz for their valuable feedback
and Yufeng Zhou for providing us with VMProtect samples.

\bibliographystyle{plainurl}
\bibliography{\jobname}
\cleardoublepage

\appendix

\section{Options for Tigress Evaluation}
\label{appendix:tigress_opts}

\subsection{Tigress Command Line}
\label{appendix:tigress_command_line}

\begin{lstlisting}[language=bash, basicstyle=\fontfamily{SourceCodePro-TLF}\selectfont\tiny,float, caption={Tigress options used to
compile samples without super operators.}, label={listing:tigress_opts}]
./tigress \
  --Environment=x86_64:Linux:Gcc:4.6 \
  --Seed=42 --Transform=Virtualize \
      --Functions={functions} \
      --VirtualizeDispatch=ifnest \
      --VirtualizeOperands={operandmode} \
  --out={obf_output} {input}
gcc {obf_output} -o {output} -gdwarf-4
\end{lstlisting}

\begin{lstlisting}[language=bash, basicstyle=\fontfamily{SourceCodePro-TLF}\selectfont\tiny,float, caption={Tigress options used to
compile samples with super operators.}, label={listing:tigress_opts_sopts}]
./tigress \
  --Environment=x86_64:Linux:Gcc:4.6 \
  --Seed=42 --Transform=Virtualize \
      --Functions={functions} \
      --VirtualizeDispatch=ifnest \
      --VirtualizeOperands={operandmode} \
      --VirtualizeMaxDuplicateOps=2 \
      --VirtualizeMaxMergeLength=10 \
      --VirtualizeSuperOpsRatio=2.0 \
  --out={obf_output} {input}
gcc {obf_output} -o {output} -gdwarf-4
\end{lstlisting}

\subsection{Compilation options for \autoref{subsection:code_similarity}}\label{appendix:redec_opts}
\begin{lstlisting}[language=bash, basicstyle=\fontfamily{SourceCodePro-TLF}\selectfont\tiny, caption={Compile flags before x64 decompilation step}]
gcc -std=c11 -fpermissive \
    -fno-strict-aliasing \
    -fno-stack-protector
    -Wno-int-conversion \
    -fno-inline \
    -mno-sse \
    -O2 -o {output} {input} \
    {linked_libraries}
\end{lstlisting}

\section{Code Virtualizer}\label{appendix:codevirtualizer}

\subsection{Compilation options for Code Virtualizer binaries}\label{appendix:codevirtualizer_sample_opts}
\begin{lstlisting}[language=bash, basicstyle=\fontfamily{SourceCodePro-TLF}\selectfont\tiny, caption={Compile flags for Code Virtualizer obfuscation.}]
gcc -I{codevirtualizer-include-dir} \
    -fno-reorder-blocks \
    -fno-reorder-blocks-and-partition \
    -fno-tree-tail-merge \
    -fno-stack-protector \
    -fno-inline \
    -O2 -mno-red-zone \
    -o {output} {input} \
    {linked_libraries}
\end{lstlisting}

\begin{figure}[htb]
	\centering
	\caption{Comparison of output from \toolname{} on \texttt{binarysearchrec} obfuscated with Code Virtualizer Fish. Before and after recompilation.}
	\label{appendix:rq3_binarysearchrec_cv}

	\tabskip=0pt
	\begin{subfigure}{0.4\textwidth}
		\caption{\toolname{}'s output after deobfuscation.}
		\begin{lstlisting}[language=reC,basicstyle=\fontfamily{SourceCodePro-TLF}\selectfont\tiny]
undefined8
binary(longlong param_1, undefined8 param_2, undefined8 param_3, longlong param_4, longlong param_5)
{
  ulonglong uVar1;
  ulonglong uVar2;
  int iVar3;
  int iVar4;
  undefined8 uVar5;
  
  iVar3 = (int)param_4;
  iVar4 = iVar3 - (int)param_5;
  if (iVar4 < 0 == (iVar3 < 0 != (int)param_5 < 0 && iVar3 < 0 != iVar4 < 0)) {
    return 0;
  }
  uVar1 = (longlong)((int)((param_4 + param_5 & 0xffffffffUL) >> 0x1f) + (int)(param_4 + param_5))
          >> 1;
  uVar2 = uVar1 & 0xffffffff;
  iVar3 = *(int *)(((ulonglong)(uint)((int)uVar1 >> 0x1f) << 0x20 ^ uVar2) * 4 + param_1);
  iVar4 = (int)param_3;
  if (iVar3 == iVar4) {
    return 1;
  }
  if ((iVar3 != iVar4) &&
     (iVar3 - iVar4 < 0 == (iVar3 < 0 != iVar4 < 0 && iVar3 < 0 != iVar3 - iVar4 < 0))) {
    uVar5 = binary(param_1,param_2,param_3,param_4,uVar2 - 1 & 0xffffffff,uVar2);
    return uVar5;
  }
  uVar5 = binary(param_1,param_2,param_3,uVar2 + 1 & 0xffffffff,param_5,uVar2);
  return uVar5;
}

		\end{lstlisting}
	\end{subfigure}
	\begin{subfigure}{0.4\textwidth}
		\caption{\toolname{}'s output after deobfuscation.}
		\caption{Ghidra's output after decompiling the deobfuscated (and optimized) binary.}
		\begin{lstlisting}[language=reC,basicstyle=\fontfamily{SourceCodePro-TLF}\selectfont\tiny]
undefined8 binary(long param_1,undefined8 param_2,uint param_3,uint param_4,uint param_5)
{
  uint uVar1;
  int iVar2;
  int iVar3;
  
  while( true ) {
    if (-1 < (int)((param_4 ^ param_5) & (param_4 - param_5 ^ param_4) ^ param_4 - param_5)) {
      return 0;
    }
    iVar2 = (param_4 + param_5) - ((int)(param_4 + param_5) >> 0x1f);
    iVar3 = iVar2 >> 1;
    uVar1 = *(uint *)(param_1 + CONCAT44(iVar2 >> 0x1f,iVar3) * 4);
    if (uVar1 == param_3) break;
    if ((int)((uVar1 ^ uVar1 - param_3) & (uVar1 ^ param_3) ^ uVar1 - param_3) < 0) {
      param_4 = iVar3 + 1;
    }
    else {
      param_5 = iVar3 - 1;
    }
  }
  return 1;
}



		\end{lstlisting}
	\end{subfigure}
\end{figure}

\vspace{30em}

\end{document}